\documentclass[a4paper,11pt]{article}
\pdfoutput=1 % if your are submitting a pdflatex (i.e. if you have
\usepackage{jheppub} % for details on the use of the package, please
\usepackage[T1]{fontenc} % if needed
\usepackage{xcolor} % Add this line
\usepackage{hyperref} % You can still use hyperref for other options

\usepackage{booktabs}

\usepackage{multirow} 

\usepackage{tabularx} 

\usepackage{makecell}

\title{\boldmath Dual Inflation and Bounce Cosmologies Interpretation of Pulsar Timing Array Data}

\author[a]{Changhong Li\footnote{Corresponding Author}}

\author[a]{Junrong Lai}
\author[a]{Jinjie Xiang}
\author[a]{Chaofan Wu}
\affiliation[a]{Department of Astronomy, Key Laboratory of Astroparticle Physics of Yunnan Province,\\  School of Physics and Astronomy,  Yunnan University,\\ No.2 Cuihu North Road, Kunming,  650091 China}
\emailAdd{changhongli@ynu.edu.cn}
\emailAdd{clare5251ljr@gmail.com}
\emailAdd{1770152009@qq.com}
\emailAdd{1271074168@qq.com}

\abstract{We explore a dual scenario of generalized inflation and bounce cosmologies, producing a scale-invariant curvature perturbation spectrum. Bayesian analysis with pulsar timing array data identifies, for the first time, viable regions from inflation and bounce that simultaneously explain stochastic gravitational wave background (SGWB) signals and CMB anisotropies. Bayes factor calculations strongly favor this dual scenario over conventional sources and provide initial evidence of a duality between inflation and bounce regarding SGWB, offering new insights for early universe model-building and future observations.}

\begin{document}
\maketitle
\flushbottom

%%%%%%%%%%%%%%%%%%%%%%%%%%%%%%%%%%%%%%%%%%%%%%%%%%%%%
\section{Introduction}    \label{sec:intro}  

Pulsar Timing Array (PTA) experiments, including NANOGrav \cite{NANOGrav:2020spf, NANOGrav:2023gor}, EPTA \cite{EPTA:2021crs, EPTA:2023fyk}, PPTA \cite{Goncharov:2021oub, Reardon:2023gzh}, IPTA \cite{Antoniadis:2022pcn}, and CPTA \cite{Xu:2023wog}, provide a unique probe into the new physics related to nano-Hertz frequency gravitational waves \cite{Caprini:2018mtu}. Various astrophysical and cosmological sources, including supermassive black hole binaries (SMBH binaries)\cite{Sesana:2013wja, Kelley:2017lek, Chen:2018znx, Burke-Spolaor:2018bvk}, inflationary gravitational waves (IGWs)\cite{Zhao:2013bba, Guzzetti:2016mkm, Vagnozzi:2020gtf}, cosmic strings (CS) \cite{Siemens:2006yp, Cui:2018rwi, Gouttenoire:2019kij}, domain walls (DWs) \cite{Hiramatsu:2013qaa, Ferreira:2022zzo,Bian:2022qbh}, first-order phase transitions (FOPT) \cite{Hindmarsh:2013xza, Hindmarsh:2015qta, Caprini:2015zlo, Hindmarsh:2017gnf, Gouttenoire:2023bqy, Salvio:2023ynn}, and scalar-induced gravitational waves (SIGWs) \cite{Ananda:2006af, Baumann:2007zm, Kohri:2018awv}, have been proposed to explain the recently detected stochastic gravitational wave background (SGWB) signals \cite{NANOGrav:2023hvm, Figueroa:2023zhu}. However, some Bayesian analyses suggest that no single source is strongly preferred over the others (see, for example, \cite{Bian:2023dnv, Ellis:2023oxs}). To further explore the physical implications of SGWB signals, various models have been proposed, such as audible axion \cite{Machado:2019xuc, Machado:2018nqk}, kination-domination \cite{Co:2021lkc, Oikonomou:2023qfz}, preheating \cite{Bethke:2013aba, Adshead:2019igv}, and particle production \cite{Dimastrogiovanni:2016fuu, DAmico:2021fhz}. For a comprehensive review, see \cite{Caprini:2018mtu}, and for recent developments, see \cite{NANOGrav:2023hvm, Figueroa:2023zhu, Bian:2023dnv, Ellis:2023oxs} and references therein.

Among these SGWB candidates, inflationary gravitational waves (IGWs) with a blue-tilted tensor spectrum, $n_T \simeq 1.8 \pm 0.3$ \cite{Vagnozzi:2023lwo}, are compelling. However, constrained by the ``consistency'' relations $n_T \simeq 1-n_s$ and $n_T = -r/8$ \cite{Figueroa:2023zhu, Vagnozzi:2023lwo}, this interpretation challenges standard slow-roll inflation. Here, $n_T$ and $1-n_s$ denote the spectral indices of tensor and scalar (curvature) perturbations, respectively, and $r$ represents the tensor-to-scalar ratio. This conflict indicates that standard inflation cannot concurrently explain CMB anisotropies ($n_s-1=-0.04$ \cite{WMAP:2010qai, Planck:2015fie, Planck:2018vyg}) and the SGWB signals detected by PTAs, as conventionally, a scale-invariant curvature perturbation spectrum cannot coexist with a blue-tilted primordial gravitational wave spectrum \cite{Figueroa:2023zhu}. Similar challenges are evident in other inflation and bounce models \cite{Brandenberger:2006xi, Brandenberger:2014faa, Brandenberger:2009yt, Cai:2011zx, Khoury:2001wf, Cabass:2015jwe, Wang:2016tbj, Calcagni:2020tvw, Li:2022uip,Choudhury:2023jlt, Choudhury:2023vuj}.

Notably, a dual scenario combining generalized inflation and bounce cosmologies can potentially address this challenge \cite{Li:2013bha}. By introducing a phenomenologically generalized redshift term for curvature perturbations, this scenario maintains scale invariance with $(n_s-1) \simeq -0.04$ despite a blue-tilted tensor spectrum of $n_T = 1.8 \pm 0.3$. Extending Wands's duality~\cite{Wands:1998yp, Finelli:2001sr, Boyle:2004gv, Raveendran:2023auh}, this approach extends the standard Hubble horizon, $\mathcal{H}^{-1} \equiv (aH)^{-1}$, to an ``effective'' Hubble horizon, $\mathcal{H}_u^{-1} \equiv |aH|^{-1}$, incorporating the collapsing phase of bounce cosmologies ($aH<0$) \cite{Li:2013bha}. Additionally, this scenario generalizes the redshift term coefficient of curvature perturbation modes $\chi_k$ from $3H$ to $mH$, leading to~\cite{Li:2013bha}:
\begin{equation}
     \ddot{\chi}_k + m H \dot{\chi}_k + \frac{k^2}{a^2} \chi_k = 0~,\quad  a(\eta) \propto \eta^\nu.
\end{equation}
Here, $k$ represents the wave-vector, and $H$ denotes the Hubble parameter. The parameter $\nu$ signifies the power-law index of the scale factor $a(\eta)$ in terms of conformal time $\eta$, characterizing cosmic background evolution during the perturbation horizon-exiting phase. The parameter $m$ serves as the modified damping parameter, varying with the specific generalization of inflation and bounce cosmology models. For instance, in the standard inflation model, $m=3$, while in the coupled scalar-tachyon bounce (CSTB) model, $m=0$, as the redshift term is coupled to an exotic condensation field \cite{Li:2013bha, Li:2011nj, Zhang:2019tct}. These generalizations unify various inflation and bounce models into a single parameter space $(m, \nu)$ and generalize the spectral index of scalar (curvature) perturbations, $n_s - 1 = -|(m-1)\nu-1| + 3$ (see Eq.~(\ref{eq:nsmv})) \cite{Li:2013bha}, distinguishing it from the spectral index of tensor modes, $n_T = 2\nu + 2$. This approach allows for examining the entire parameter space to find models with appropriate $(m, \nu)$ values capable of explaining both the SGWB signal and CMB anisotropies.

In this work, we explore the novel possibility that SGWB signals originate from this dual scenario combining generalized inflation and bounce cosmologies. We investigate its entire parameter space to identify regions capable of explaining both SGWB signals and CMB anisotropies. Our methodology includes imposing the scale-invariance condition for the curvature perturbation spectrum and performing Bayesian analyses using data from NANOGrav 15-year (NG15)~\cite{ZenodoNG, NANOGrav:2023hde, NANOGrav:2023gor, NANOGrav:2023hvm}, EPTA DR2~\cite{ZenodoEPTA, EPTA:2023sfo, EPTA:2023fyk}, PPTA DR3~\cite{PPTADR3, Zic:2023gta, Reardon:2023gzh}, and IPTA DR2~\cite{IPTADR2, Perera:2019sca, Antoniadis:2022pcn} datasets (hereafter PTAs' datasets).

For the first time, we identify two groups of four regions within this dual scenario that simultaneously explain SGWB signals and CMB anisotropies. Our methodology involves three steps: first, within the conventional parameter space $(n_T, r)$, we perform Bayesian analysis with PTA data to identify regions for the dual scenario that explain SGWB signals (Fig.~\ref{fig:Triunfold.pdf}a), consistent with previous studies \cite{Vagnozzi:2023lwo}. Second, within the parameter space $(w, r)$, using the relation between $n_T$ and $w$, we perform Bayesian analysis with PTA data again to isolate inflation and bounce regions (Fig.~\ref{fig:Triunfold.pdf}b). Finally, by enforcing scale invariance for the curvature perturbation spectrum and conducting further Bayesian analyses with PTA data, we for the first time delineate two groups of four regions that account for SGWB signals and CMB anisotropies in the $(m, \nu)$ parameter space (Fig.~\ref{fig:Triunfold.pdf}c). These groups correspond to stable and dynamic scale-invariant solutions, each comprising dual regions from inflation and bounce cosmologies.

\begin{figure}[htbp]
\centering 
\includegraphics[width=0.9\textwidth]{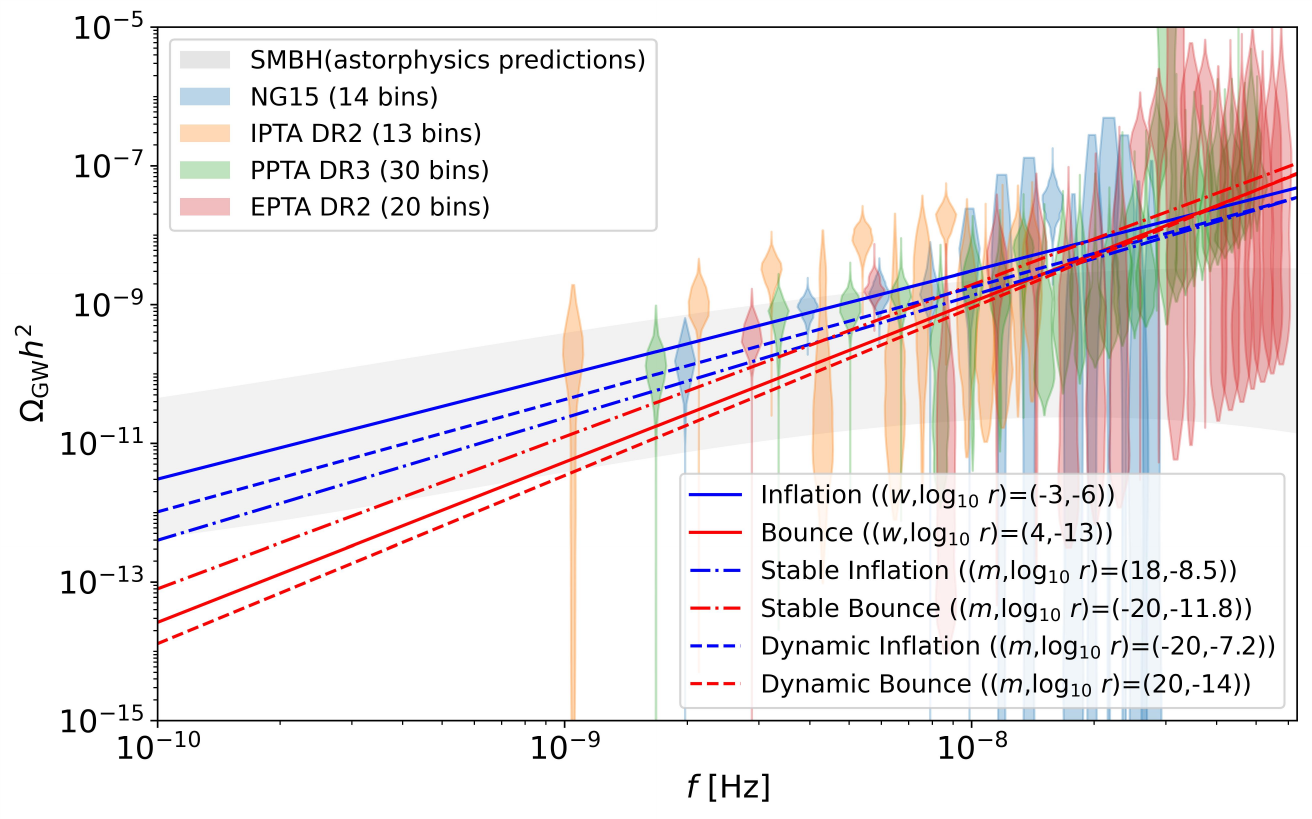}
\caption{\label{fig:violin6.pdf} Examples from the $1~\sigma$ posterior distribution region of our findings within the parameter spaces $(w,r)$ and $(m,r)$ (see Fig.\ref{fig:Triunfold.pdf}~(b) and (c)), illustrating their potential to explain the SGWB signals detected by PTAs. The "violin" diagrams depict the posterior probability distribution of the SGWB spectrum across various frequency bins in these PTAs' datasets \cite{NANOGrav:2023gor,EPTA:2023fyk, Reardon:2023gzh, Antoniadis:2022pcn, ZenodoNG, NANOGrav:2023hde, ZenodoEPTA, EPTA:2023sfo, PPTADR3, Zic:2023gta, IPTADR2, Perera:2019sca}, plotted using code from \cite{Gouttenoire:2023bqy}. The shaded gray regions represent astrophysical predictions from supermassive black hole (SMBH) binaries \cite{Rosado:2015epa}.}
\end{figure}

In Fig.~\ref{fig:violin6.pdf}, we present two examples from the $1\sigma$ posterior distribution region within the parameter spaces $(w,r)$ (see Fig.~\ref{fig:Triunfold.pdf}(b)) and four examples from $(m,r)$ (see Fig.~\ref{fig:Triunfold.pdf}(c)), demonstrating their potential to explain the SGWB signals within PTAs' datasets \cite{NANOGrav:2023gor, EPTA:2023fyk, Reardon:2023gzh, Antoniadis:2022pcn, ZenodoNG, NANOGrav:2023hde, ZenodoEPTA, EPTA:2023sfo, PPTADR3, Zic:2023gta, IPTADR2, Perera:2019sca}. According to our findings, the latter four examples can simultaneously address both SGWB signals and CMB anisotropies.

To assess the credibility of this novel scenario against other astrophysical and cosmological sources, we compute the Bayes factors (BFs) between them. Our analysis reveals robust support for this dual scenario from the NANOGrav 15-year data, with significantly higher BFs compared to conventional sources such as supermassive black hole binaries, cosmic strings, domain walls, first-order phase transitions, and scalar-induced gravitational waves, as listed in Table~\ref{Tab:BFMCNG15}. Moreover, BFs computed within the dual scenario regions, all close to unity, suggest initial evidence of a duality between inflation and bounce cosmologies regarding SGWB, as shown in Table~\ref{Tab:dualinbo}. These findings underscore the importance of further investigation into this dual scenario and its implications for future model-building in the early universe's inflation or bounce phases. Additionally, we highlight parameter regions to be explored by future CMB observations and gravitational wave detectors, noting some regions already excluded by recent observations.

\section{Dual Scenario Interpretation of SGWB Spectrum}
The SGWB spectrum induced by primordial gravitational waves from inflation and bounce cosmologies is given by \cite{Caprini:2018mtu}: 
\begin{equation}
     \Omega_\mathrm{GW}(f)h^2=\frac{3}{128}\Omega_{\gamma 0}h^2\cdot r\cdot\mathcal{P}_\mathcal{R}\left(\frac{f}{f_\ast}\right)^{n_T}\left[\frac{1}{2}\left(\frac{f_\mathrm{eq}}{f}\right)^2+\frac{16}{9}\right]~, \label{eq:sgwbinrnt}
\end{equation}
where $r$ and $n_T$ are the tensor-to-scalar ratio ($r\equiv \mathcal{P}_h/\mathcal{P}_\mathcal{R}$) and the spectral index of the primordial tensor spectrum ($\mathcal{P}_h\propto (k/k_\ast)^{n_T}$), measured at the pivot scale $k_\ast = 0.05\mathrm{Mpc}^{-1}$. $\mathcal{P}_\mathcal{R} = 2 \times 10^{-9}$ is the amplitude of the curvature perturbation spectrum at $k=k_\ast$ \cite{Caprini:2018mtu}. $f_\ast = 1.55\times 10^{-15}~\mathrm{Hz}$ is the frequency today corresponding to $k_\ast$, and $f_\mathrm{eq} = 2.01\times 10^{-17}~\mathrm{Hz}$ is the frequency today corresponding to matter-radiation equality. $\Omega_{\gamma 0} = 2.474 \times 10^{-5} h^{-2}$ is the energy density fraction of radiation today, and $h = 0.677$ is the reduced Hubble constant.

\begin{figure}
\centering 
\includegraphics[width=1.0\textwidth]{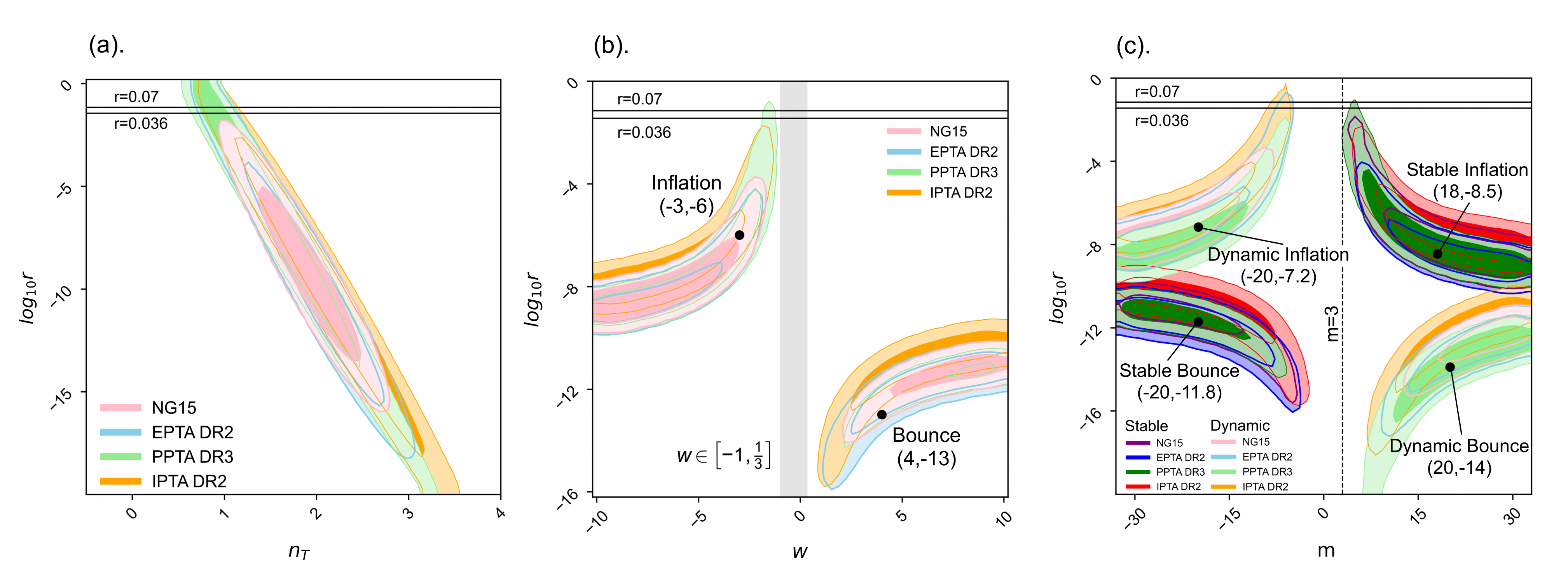} 
\caption{\label{fig:Triunfold.pdf} Posterior distributions (at $68\%$ and $95\%$ confidence levels) in the parameter spaces $(n_T,r)$, $(w,r)$, and $(m,r)$ for the dual scenario of generalized inflation and bounce cosmologies. Bayesian analyses across PTA datasets employed three different approaches: direct fitting of Eq.(\ref{eq:sgwbinrnt}), fitting Eq.(\ref{eq:sgwbinrnt}) with Eq.(\ref{eq:ntwexp}), and fitting Eq.(\ref{eq:sgwbinrnt}) with Eq.(\ref{eq:mnustacon}) and Eq.(\ref{eq:mnuunstacon}). Sub-figure (a) depicts the viable region  of this dual scenario within the conventional parameter space $(n_T,r)$ explaining the SGWB signal. Sub-figure (b) delineates isolated regions for inflation and bounce cosmologies within $(w,r)$ parameter space capable of explaining the SGWB signal. Sub-figure (c) reveals two groups of four regions in $(m,r)$ parameter space, newly identified in this dual scenario, capable of explaining SGWB signals and CMB anisotropies concurrently. These sub-figures also illustrate the gradual resolution of degeneracy resulting from the duality between inflation and bounce, and the balance between the damping effect of perturbations ($m$) and the temporal effect of the background ($\nu$) in the parameter space.}
\end{figure}

Initially, we performed a Bayesian analysis to fit Eq.~(\ref{eq:sgwbinrnt}) using PTAs' datasets \cite{NANOGrav:2023gor, EPTA:2023fyk, Reardon:2023gzh, Antoniadis:2022pcn, ZenodoNG, NANOGrav:2023hde, ZenodoEPTA, EPTA:2023sfo, PPTADR3, Zic:2023gta, IPTADR2, Perera:2019sca}, yielding the 2D joint posterior probability distribution in the conventional parameter space $(n_T, r)$ (Fig.~\ref{fig:Triunfold.pdf}(a)). This distribution aligns with previous studies (e.g., Fig.~2 in Ref.~\cite{Vagnozzi:2023lwo}). Notably, this dual scenario interprets the viable range of $n_T$ from both inflation ($n_T = \frac{4}{3w+1} + 2 < 2$ for $w < -\frac{1}{3}$) and bounce cosmologies ($n_T = \frac{4}{3w+1} + 2 > 2$ for $w > -\frac{1}{3}$), as shown in Eq.~(\ref{eq:ntwexp}). Detailed information on the Bayesian analysis is provided in the Appendix.

Subsequently, to resolve the degeneracy between inflation and bounce cosmologies, we express the tensor spectral index $n_T$ in terms of the equation of state $w$:
\begin{equation}
    n_T = \frac{4}{3w+1} + 2~. \label{eq:ntwexp}
\end{equation}
This formulation mirrors the conventional relation in inflation cosmologies ($\dot{a} > 0$ and $\ddot{a} > 0$; $w < -\frac{1}{3}$)~\cite{Zhao:2013bba, Vagnozzi:2023lwo}, and also incorporates the contraction phase of bounce cosmologies ($\dot{a} < 0$ and $\ddot{a} < 0$; $w > -\frac{1}{3}$).

Substituting Eq.~(\ref{eq:ntwexp}) into Eq.~(\ref{eq:sgwbinrnt}) and performing a Bayesian analysis using PTAs datasets, we identified two isolated regions in the parameter space $(w,r)$ that can explain the SGWB signals (Fig.~\ref{fig:Triunfold.pdf}(b)): inflation ($w<-\frac{1}{3}$, top left) with larger $r$ values and bounce cosmologies ($w>-\frac{1}{3}$, bottom left) with smaller $r$ values. The conventional EoS region $-1 < w < \frac{1}{3}$, encompassing traditional inflation and bounce models, is not favored by this Bayesian analysis, as shown by the shaded gray region in Fig.~\ref{fig:Triunfold.pdf}(b). This suggests that standard inflation and matter bounce models \cite{Brandenberger:2009yt,Cai:2011zx} cannot account for the SGWB signal, corroborating previous findings \cite{Figueroa:2023zhu}. Our results provide the first systematic demonstration of the potential to account for SGWB signals within a broader parameter space of $w$ in inflationary and bounce cosmologies. Additionally, future experiments like CMB-S4 \cite{CMB-S4:2016ple} will explore further regions of the inflationary parameter space of $r$, beyond those ($-1.92<w<-1.18$ for $2\sigma$ C.L. range of PPTA DR3 dataset) excluded by current CMB anisotropy observations at $r=0.07$ (WMAP and Planck data) \cite{WMAP:2010qai, Planck:2015fie, Planck:2018vyg} and $r=0.036$ (BICEP+) \cite{BICEP:2021xfz, Tristram:2021tvh} .

Ultimately, by enforcing scale invariance conditions for the curvature perturbation spectrum and performing further Bayesian analyses using PTAs datasets, we identify the viable regions within the dual scenario that can concurrently account for SGWB signals and CMB anisotropies. Specifically, in this dual scenario, the spectral index of the curvature perturbation spectrum, $\mathcal{P}_\mathcal{R} \propto k^{n_s-1}$, is given by
\begin{equation}\label{eq:nsmv}
n_s-1=-|(m-1)\nu-1|+3~,
\end{equation}
(cf. Eq.~(4.2) and Eq.~(4.3) in Ref.~\cite{Li:2013bha}, also see Appendix), where $\nu$ is the power-law index of the scale factor and $m$ is the damping parameter of the generalized redshift term. Applying the scale-invariance condition, $n_s-1 = 0$ (the case of $n_s-1 = -0.04$ is detailed in the Appendix), two solutions can be derived from Eq.~(\ref{eq:nsmv}) (cf. Eq.~(4.5) and Eq.~(4.6) in Ref.~\cite{Li:2013bha}): a time-independent (stable) scale-invariant solution ($\mathcal{P}_\mathcal{R}\propto k^0\eta^0$),
\begin{equation}\label{eq:mnustacon}
m = -\frac{2}{\nu} + 1~,
\end{equation}
and a time-dependent (dynamic) scale-invariant solution ($\mathcal{P}_\mathcal{R}\propto k^0\eta^{-2(m-1)\nu+2}$),
\begin{equation}\label{eq:mnuunstacon}
m = \frac{4}{\nu} + 1~.
\end{equation}
In these two solutions, the damping parameter $m$, which characterizes the enhanced or suppressed damping effect, balances the power index of the scale factor $\nu$, which characterizes the temporal effect from the cosmic background, to generate scale-invariant curvature perturbation spectra compatible with CMB anisotropy observations. For instance, the standard inflation model $(m,\nu)=(3,-1)$ and the CSTB model $(m,\nu)=(0,2)$ correspond to the stable solution in Eq.~(\ref{eq:mnustacon}), while the matter bounce model $(m,\nu)=(3,2)$ with $\mathcal{P}_\mathcal{R}\propto k^0\eta^{-6}$ corresponds to the dynamic solution in Eq.~(\ref{eq:mnuunstacon}) \cite{Brandenberger:2009yt,Cai:2011zx}.

Utilizing the relation $n_T = 2\nu + 2$, substituting Eq.~(\ref{eq:mnustacon}) and Eq.~(\ref{eq:mnuunstacon}) into Eq.~(\ref{eq:sgwbinrnt}), and performing further Bayesian analysis using PTAs datasets, we identified, for the first time, two groups of four regions—each group comprising dual regions from inflation and bounce—that can simultaneously explain the SGWB signal and CMB anisotropies, as illustrated in Fig.~\ref{fig:Triunfold.pdf}(c). The stable and dynamic scale-invariant groups are depicted as the darker (Eq.~(\ref{eq:mnustacon})) and lighter (Eq.~(\ref{eq:mnuunstacon})) colored regions, respectively. This finding addresses the challenge faced by conventional inflation and bounce cosmologies and provides the first systematic demonstration of the potential to account for SGWB signals within a broader parameter space of $w$ in inflation and bounce cosmologies, shedding light on future early-universe model-building.

According to our findings, the parameter space around the standard damping parameter value $m=3$ has been excluded by data from PTAs' datasets \cite{NANOGrav:2023gor, EPTA:2023fyk, Reardon:2023gzh, Antoniadis:2022pcn, ZenodoNG, NANOGrav:2023hde, ZenodoEPTA, EPTA:2023sfo, PPTADR3, Zic:2023gta, IPTADR2, Perera:2019sca}, corroborating numerous previous studies~\cite{Caprini:2018mtu}. For general values of $m$, some parameter regions associated with $r$ have been ruled out by current observations of CMB anisotropies (as constrained by $r=0.036$ from BICEP+, $m$ has been excluded for $[-8.4, -4.7]$ and $[4.1, 5.7]$ at the $2\sigma$ confidence level of the PPTA DR3 dataset). However, significant unexplored territories remain, which upcoming experiments like CMB-S4 are well-positioned to investigate. The overlap of colored regions between inflation and bounce arises from uncertainties across PTAs' datasets. For each dataset, these regions do not overlap.

{\it Model Comparison.}--- In Table~\ref{Tab:BFMCNG15}, we present the Bayes factors (BF) 
\begin{equation}
    B_{\alpha j} = \mathrm{evidence}[H_\alpha]/\mathrm{evidence}[H_j]~,
\end{equation}
obtained from a Bayesian model selection analysis using the NG15 dataset. Here, $H_\alpha$ denotes each case within scenario $\alpha$ (including the Dual Scenario in $(w,r)$ prior, Stable, Dynamic, and Dual Scenario in $(m,r)$ prior), and $H_j$ refers to each model $j$ (including SMBHBs, CS, DW, FOPT, and SIGW). The results indicate that all BFs are greater than 20, suggesting that the dual scenario fits the NG15 data significantly better than other astrophysical and cosmological sources of SGWB. Note that for model comparisons, a Bayes factor $B_{ij}$ of 20 between a candidate model $\mathrm{M}_i$ and another model $\mathrm{M}_j$ corresponds to a 95\% belief in the statement ``$\mathrm{M}_i$ is true,'' indicating strong evidence in favor of $\mathrm{M}_i$ \cite{Bian:2023dnv}. This indicates the dual scenario as a compelling source and underscores its implications for future model-building in the early universe's inflation or bounce phases. In the Appendix, we provide BF matrices for all model comparisons across NG15, EPTA DR2, PPTA DR3, and IPTA DR2 with their prior ranges. While other PTAs' datasets are not as optimistic as NG15, they still provide substantial support for the dual scenario.

\begin{table}
\centering
\caption{Bayes factors (BF) comparing dual scenario with conventional source using NG15 data}\label{Tab:BFMCNG15}
\begin{tabular}{cccccc}
\toprule
Bayes factor &\textbf{SMBHBs} & \textbf{CS} & \textbf{DW} & \textbf{FOPT}&  \textbf{SIGW}\\ 
\midrule
\textbf{Dual Scenario $(w,r)$} &94.6 $\pm$ 34.9 & 461.6 $\pm$ 127.2 & 318.8 $\pm$ 73.2 & 61.4 $\pm$ 11.6 &65.2 $\pm$ 12.4 \\
\textbf{Stable Solutions} & 81.5 $\pm$ 32.3 & 397.7 $\pm$ 117.8 & 274.7 $\pm$ 67.8 & 52.9 $\pm$ 10.7 & 56.2 $\pm$ 10.7  \\
\textbf{Dynamic Solutions} & 49.4 $\pm$ 17.3 & 240.7 $\pm$ 63.2 & 166.3 $\pm$ 36.4 & 32.0 $\pm$ 6.8 & 34.0 $\pm$ 6.5  \\
\textbf{Dual Scenario $(m,r)$}  & 83.9 $\pm$ 30.9 & 409.0 $\pm$ 112.7 & 282.6 $\pm$ 64.9 & 54.4 $\pm$ 10.3 & 57.8 $\pm$ 11.0  \\
\bottomrule
\end{tabular}
\end{table}

In Table~\ref{Tab:dualinbo}, we present the Bayes factors comparing inflation to bounce, stable scale-invariant inflation to bounce, dynamic scale-invariant inflation to bounce, and stable to dynamic scale-invariant solutions. (R/L) represents the region on the right-hand side relative to the region on the left-hand side. The results, all close to 1, provide initial evidence of strong dualities between inflation and bounce models concerning SGWB—a duality previously established only for curvature perturbations~\cite{Li:2013bha, Wands:1998yp, Finelli:2001sr, Boyle:2004gv, Raveendran:2023auh}. A Bayes factor $B_{ij}$ of 1 indicates no preference between models \cite{Bian:2023dnv}. This finding strengthens the motivation for future model-building of bounce cosmologies.

\begin{table}[htbp]
\centering
\caption{Bayes Factors Comparing Each Regions of Dual Scenario Across Different PTA Datasets}\label{Tab:dualinbo}
\begin{tabular}{ccccc}
\toprule
Bayes factor & \textbf{NG15} & \textbf{EPTA} & \textbf{PPTA} & \textbf{IPTA} \\ \midrule
\textbf{Bounce/Inflation} & $0.95 \pm 0.22$ & $1.23 \pm 0.32$ & $1.48 \pm 0.61$ &  $1.37 \pm 0.36$\\
\textbf{Stable(R/L)} & 0.94 $\pm$ 0.24 & 0.89 $\pm$ 0.22 & 0.95 $\pm$ 0.43 & 1.01 $\pm$ 0.42\\
\textbf{Dynamic(R/L)} & 0.63 $\pm$ 0.16 & 1.41 $\pm$ 0.27 & 1.24$\pm$ 0.39 & 1.08 $\pm$ 0.41 \\
\textbf{Stable/Dynamic} &1.65 $\pm$ 0.46 & 1.33 $\pm$ 0.44 & 1.23$\pm$ 0.56 & 1.49 $\pm$ 0.79\\
\bottomrule
\end{tabular}
\end{table}

\section{Conclusion} 
%%%%%%%%%%%%%%%%%%%%%%%%%%%%%%%%%%%%%%%%%%%%%%%%%%%%%
We explored the entire parameter space of a dual scenario combining generalized inflation and bounce cosmologies. By performing Bayesian analysis using PTAs' datasets, we identified, for the first time, two groups of four distinct regions—each group comprising dual regions from both inflation and bounce—that can simultaneously explain SGWB signals and CMB anisotropies.

The Bayes factor calculations indicate that this dual scenario is strongly favored by NG15 data, with large BFs compared to conventional sources such as SMBHBs, CS, DW, FOPT, and SIGW. This finding suggests the dual scenario is compelling and offers new insights for future model-building of inflation and bounce cosmologies. Additionally, the BFs between each region of the dual scenario, all close to unity, suggest a duality between inflation and bounce regarding SGWB, previously established only for curvature perturbations. This strengthens the motivation for bounce cosmology.

Beyond this initial exploration, further investigations should focus on the impact of the reheating and bounce process \cite{Caprini:2018mtu}, constraints imposed by $\Delta N_\mathrm{eff}$~\cite{NANOGrav:2023hvm, Figueroa:2023zhu}, implications from CMB spectral distortion~\cite{Tagliazucchi:2023dai, Ramberg:2022irf, Chluba:2012we, Jeong:2014gna, Nakama:2014vla}, and the need for a broken power law of SGWB spectra ~\cite{Kuroyanagi:2020sfw, Benetti:2021uea} to accommodate LIGO/Virgo/KAGRA and other observations \cite{LIGOScientific:2014pky, VIRGO:2014yos, KAGRA:2018plz, LISA:2017pwj, Kawamura:2011zz, Punturo:2010zz, Nan:2011um, Ruan:2018tsw, TianQin:2015yph} at high frequencies. Although in some parameter spaces the value of $r$ predicted by this scenario, especially for bounce cosmology ($r < 10^{-10}$), is too small to be detected, the predicted SGWB signals and the broken scale of the power law can still be explored at higher frequencies by these observatories. Many constructions of cosmological inflation and bounce models, especially those aimed at explaining SGWB signals, are approaching the viable region identified in this work (see \cite{Lehners:2008vx, Cai:2014bea, Choudhury:2023hfm, Choudhury:2023kam, Choudhury:2023hvf, Calcagni:2020tvw, Choudhury:2024dei, Papanikolaou:2024fzf, Gasperini:2016gre, Ben-Dayan:2018ksd, Hu:2023ndc, Jiang:2023qht, Li:2019jwh, Brandenberger:2020tcr} for examples). Further studies relating these models and their variants to the dual scenario are warranted.

\acknowledgments
We would like to thank Yann Gouttenoire and Junchao Zong for their valuable communications on data analysis, and Jing Shu for useful discussions. C.L. is supported by the NSFC under Grants No.11963005 and No. 11603018, by Yunnan Provincial Foundation under Grants No.202401AT070459, No.2019FY003005, and No.2016FD006, by Young and Middle-aged Academic and Technical Leaders in Yunnan Province Program, by Yunnan Provincial High level Talent Training Support Plan Youth Top Program, by Yunnan University Donglu Talent Young Scholar, and by the NSFC under Grant No.11847301 and by the Fundamental Research Funds for the Central Universities under Grant No. 2019CDJDWL0005.

%%%%%%%%%%%%%%%%%%%%%%%%%%%%%%%%%%%%%%%%%%%%%%%%%%%%%
\appendix

\section{Dual Scenario of Generalized Inflation and Bounce Cosmologies}
This section provides a concise introduction to the dual scenario of generalized inflation and bounce cosmologies. The calculations of curvature perturbations follow the methodology outlined in Ref.~\cite{Li:2013bha}.

The standard single-field slow-roll inflation is renowned for resolving early universe challenges like the flatness and horizon problems \cite{Guth:1980zm, Starobinsky:1980te, Sato:1980yn, Linde:1981mu, Albrecht:1982wi, Mukhanov:1990me}, generating a nearly scale-invariant primordial curvature perturbation spectrum consistent with current cosmic microwave background (CMB) observations \cite{WMAP:2010qai, Planck:2015fie, Planck:2018vyg}. However, it faces the initial singularity problem, which undermines its physical basis \cite{Borde:1993xh}. Bounce cosmology offers a compelling alternative by naturally circumventing this singularity, starting with a contracting phase before transitioning to expansion through a bounce process \cite{Khoury:2001wf, Gasperini:2002bn, Creminelli:2006xe, Peter:2006hx, Cai:2007qw, Cai:2008qw, Saidov:2010wx, Li:2011nj, Cai:2011tc, Easson:2011zy, Bhattacharya:2013ut, Qiu:2015nha, Cai:2016hea, Barrow:2017yqt, deHaro:2017yll, Ijjas:2018qbo, Boruah:2018pvq, Nojiri:2019yzg}. Bounce models can also produce a nearly scale-invariant primordial curvature perturbation spectrum and may address concerns such as dark matter abundance \cite{Li:2014era, Cheung:2014nxi, Li:2014cba, Li:2015egy} and the small-scale crisis \cite{Li:2020nah}. Thus, both inflation and bounce models warrant further investigation and development in cosmology and astrophysics; see \cite{Novello:2008ra, Brandenberger:2016vhg, Nojiri:2017ncd, Odintsov:2023weg} for a comprehensive review.

\begin{figure}[htbp]
\centering 
\includegraphics[width=1.0\textwidth]{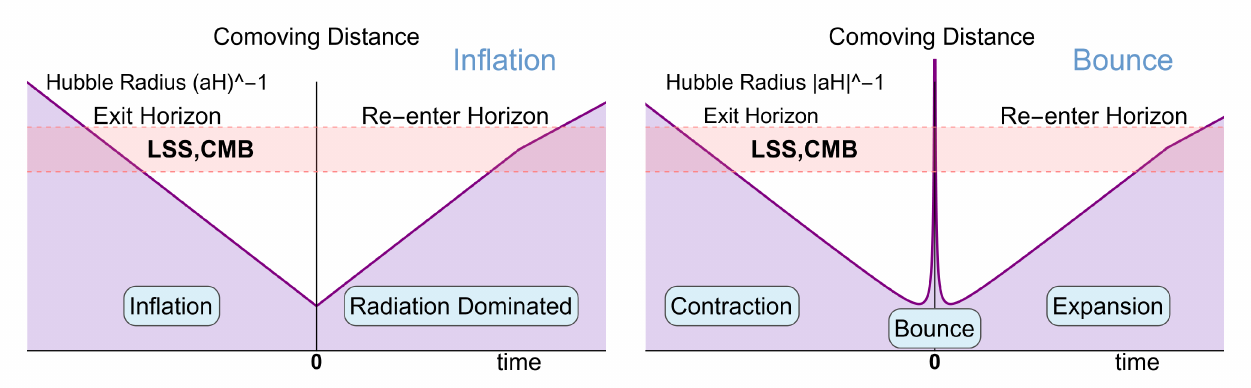}
\caption{\label{fig:inflationbounce.pdf}Left: Evolution of the Hubble radius ($(aH)^{-1}$) in the inflationary scenario. Curvature perturbation modes (horizontal pink region) exit the horizon (solid purple lines) during accelerating expansion and re-enter during decelerating expansion. Right: Evolution of the "effective" Hubble radius ($|aH|^{-1}$) in the bounce scenario, resembling the inflationary scenario. Modes exit during collapsing contraction, transition through the bounce region, and re-enter during decelerating expansion, cf. Fig. 1 in Ref.~\cite{Cheung:2016vze}.}
\end{figure}

\begin{figure}[htbp]
\centering 
\includegraphics[width=0.60\textwidth]{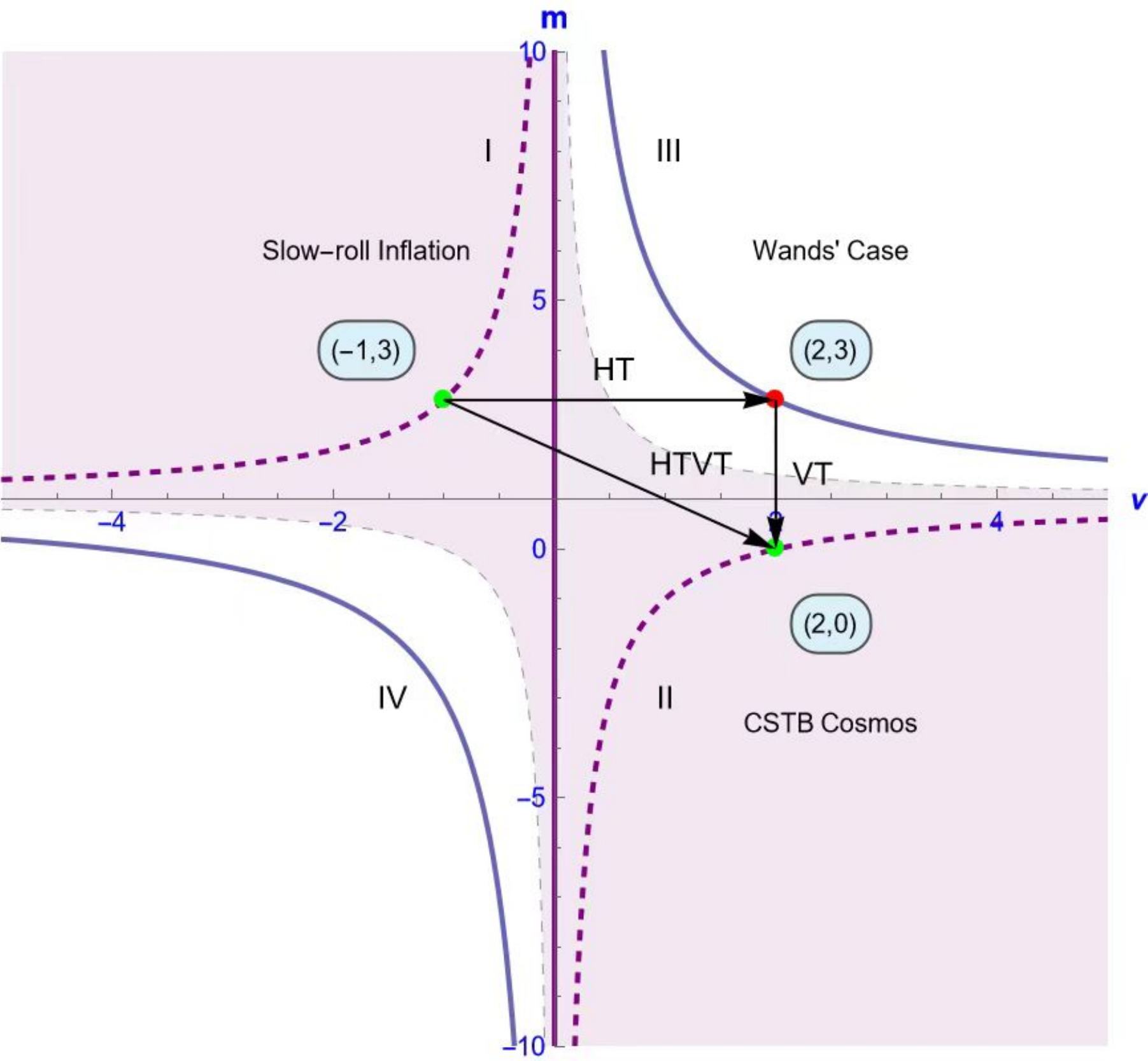}
\caption{\label{fig:duality.pdf} The parameter space $(m,\nu)$ for the scenario of generalized inflation and bounce cosmologies. The shaded region represents the range within which the time-independent (stable) curvature perturbation spectrum can be generated. The dashed curves represent the stable scale-invariant solutions for inflation (top left) and bounce (bottom right) cosmologies, while the solid curves represent the dynamic scale-invariant solutions for inflation (bottom left) and bounce (top right), cf. Fig. 2 in Ref.~\cite{Li:2013bha}.}
\end{figure}

Despite the disparate cosmic evolution trajectories of inflationary and bouncing universe models (each phase is summarized in Table~\ref{Tab:smepinbu}), they exhibit strikingly similar patterns in generating curvature perturbations, as depicted in Fig.~\ref{fig:inflationbounce.pdf} (cf. Fig. 1 in Ref.~\cite{Cheung:2016vze}). This observation prompted their integration into a unified parameter space, a concept pioneered by David Wands in Ref.~\cite{Wands:1998yp} and by Fabio Finelli and Robert Brandenberger in Ref.~\cite{Finelli:2001sr}, recognized as Wands's duality. In particular, Wands's duality is symbolized by the horizontal HT arrow in Fig.~\ref{fig:duality.pdf} (cf. Fig. 2 in Ref.~\cite{Li:2013bha}). This duality is established by extending the conventional Hubble radius $\mathcal{H}^{-1} \equiv (aH)^{-1}$, used only for inflation, to the "effective" Hubble radius $\mathcal{H}^{-1}_u \equiv |aH|^{-1}$, where $a$ and $H$ are the scale factor and Hubble parameter, respectively.

\begin{table}[ht]
    \centering
    \caption{Each phase in inflation and bounce cosmologies}\label{Tab:smepinbu}
    \begin{tabular}{llllllll}
    \toprule[0.4mm]
       & \makecell[c]{Case} &$ \makecell[c]{~\dot{a}} $& $\makecell[c]{~\ddot{a}}$ & $\makecell[c]{~\omega}$ & $\makecell[c]{~\nu}$ & $\makecell[c]{~k\eta}$ & \makecell[c]{Example} \\ \hline
       \multirow{2}{*}{Inflation} & Accelerating expansion & ~$>0$ & ~$>0$ & ~$<-1/3$ & ~$<0$ & ~$\to 0$ & ~\makecell[l]{Slow-roll inflation} \\ 
        &Decelerating expansion & ~$>0$ & ~$<0$ & ~$>-1/3$ & ~$>0$ & ~$\to \infty$ & ~\makecell[l]{Post-BBN universe} \\ 
        \hline
       \multirow{4}{*}{Bounce} &   Collapsing contraction & ~$<0$ & ~$<0$ & ~$>-1/3$ & ~$>0$ & ~$\to 0$ & ~\makecell[l]{Matter dominated} \\
       & Bouncing contraction & ~$<0$ & ~$>0$ & ~$<-1/3$ & ~$<0$ & ~$\to \infty$ & ~\makecell[l]{Quintom dominated \cite{Cai:2009zp}} \\
        & Bouncing expansion & ~$>0$ & ~$>0$ & ~$<-1/3$ & ~$<0$ & ~$\to 0$ & ~\makecell[l]{Quintom dominated \cite{Cai:2009zp}} \\
         & Decelerating expansion & ~$>0$ & ~$<0$ & ~$>-1/3$ & ~$>0$ & ~$\to \infty$ & ~\makecell[l]{Post-BBN universe}  
        \\ \toprule[0.4mm]
    \end{tabular}
\end{table}

Wands's duality provides a theoretical foundation for model-building in bounce cosmologies to generate a scale-invariant curvature perturbation spectrum, as seen in successful matter bounce models~\cite{Brandenberger:2016vhg, Nojiri:2017ncd}. However, the scale-invariant curvature spectrum from Wands's duality is time-dependent, $\mathcal{P}_\mathcal{R} \propto k^0\eta^{-4\nu+2}$. For instance, in the matter bounce model ($\nu=2$), the spectrum is $\mathcal{P}_\mathcal{R} \propto k^0\eta^{-6}$, which does not perfectly correspond to the time-independent case for inflation, $\mathcal{P}_\mathcal{R} \propto k^0\eta^0$. Here, $\mathcal{P}_\mathcal{R}$ is the power spectrum of curvature perturbations, $k$ is the wave-vector of curvature perturbation modes, $\eta$ is the conformal time, and $\nu$ is the power-law index of the scale factor, $a(\eta) \propto \eta^\nu$.

To find a time-independent scale-invariant solution in bounce cosmologies, akin to standard inflation, and to accommodate various generalized models of inflation and bounce cosmologies, the redshift term coefficient of curvature perturbation modes $\chi_k$ was phenomenologically generalized from $3H$ to $mH$, leading to (cf. Eq.~(4.1) in Ref.~\cite{Li:2013bha}):
\begin{equation}\label{eq:smchieom}
\ddot{\chi}_k + m H \dot{\chi}_k + \frac{k^2}{a^2} \chi_k = 0~,\quad a(\eta) \propto \eta^\nu~,
\end{equation}
where $m$ is assumed to be a constant that varies with the specific generalization of inflation and bounce cosmology models, characterizing various levels of damping behavior, from enhanced to suppressed effects. For instance, $m=3$ for standard inflation, whereas $m=0$ for the coupled scalar tachyon bounce (CSTB) universe model.

Due to the extension of the Hubble radius and the phenomenological generalization of the redshift term for curvature perturbations, various inflation and bounce cosmologies are unified into a single parameter space, $(\nu, m)$, as illustrated in Fig.~\ref{fig:duality.pdf}, offering new insights for the investigation of the early universe and the model-building of inflation or bounce phases. Specifically, in Fig.~\ref{fig:duality.pdf}, the shaded region represents time-independent (stable) solutions. The horizontal transformation (HT), such as Wands's duality, denotes iso-damping duality with the same $m$, while the vertical transformation (VT) represents iso-temporal duality with the same $\nu$. The combination of these transformations, HTVT, is a duality connecting stable scale-invariant solutions from inflationary and bouncing cosmologies, such as standard inflation and CSTB cosmologies.

Within the parameter space $(m,\nu)$, by solving Eq.~(\ref{eq:smchieom}), one can obtain the leading-order power spectrum of primordial curvature perturbations outside the ``effective'' horizon, $k\eta \rightarrow 0$, as (cf. Eq.~(4.2) in Ref.~\cite{Li:2013bha}):
\begin{equation}\label{eq:smprlw}
\mathcal{P}_{\mathcal{R}}\equiv \frac{k^3|\chi_k|^2}{2\pi^2}\propto k^{2L(m,\nu)+3}\eta^{2W(m,\nu)},
\end{equation}
where the indices $L$ and $W$ are functions of $m$ and $\nu$:
\begin{equation}\label{eq:smlwexp}
L(m,\nu)=-\frac{1}{2}|(m-1)\nu-1|,\quad W(m,\nu)=-\frac{1}{2}[(m-1)\nu-1+|(m-1)\nu-1|].
\end{equation}
Using Eq.~(\ref{eq:smprlw}) and Eq.~(\ref{eq:smlwexp}), one can identify the time-independent region, $2W(m,\nu)=0$, within the $(\nu, m)$ parameter space:
\begin{equation}
(m-1)\nu-1<0,
\end{equation}
as illustrated by the shaded region in Fig.~\ref{fig:duality.pdf} (cf. Fig. 3 in Ref.~\cite{Li:2013bha}). Furthermore, according to Eq.~(\ref{eq:smprlw}), the power spectrum index of curvature perturbation, $\mathcal{P}_\mathcal{R}\propto k^{n_s-1}$, is given by $n_s-1=2L(m,\nu)+3$. Using the scale-invariance condition, $n_s-1=0$, we have:
\begin{equation}\label{eq:smnslzero}
n_s-1=2L(m,\nu)+3=0.
\end{equation}
Solving Eq.~(\ref{eq:smnslzero}), we obtain two groups of scale-invariant solutions: one is the time-independent (stable) scale-invariant solution, depicted as dashed curves within the time-independent region in Fig.~\ref{fig:duality.pdf}:
\begin{equation}\label{eq:smmnurest}
m=-\frac{2}{\nu}+1,
\end{equation}
and the other is the time-dependent (dynamic) scale-invariant solution, depicted as solid curves outside the time-independent region in Fig.~\ref{fig:duality.pdf}:
\begin{equation}\label{eq:smmnureun}
m=\frac{4}{\nu}+1.
\end{equation}
In these two solutions, the damping parameter $m$, which characterizes the enhanced or suppressed damping effect, balances the power index of the scale factor $\nu$, which characterizes the temporal effect from the cosmic background, to generate scale-invariant curvature perturbation spectra. As illustrated in Fig.~\ref{fig:duality.pdf}, the models that can generate a scale-invariant curvature perturbation spectrum, including standard inflation, Wands's case (matter bounce), and CSTB cosmos, are located on these two groups of solutions.

Primordial Gravitational Waves -- In the dual scenario, the coefficient of the redshift term for tensor modes ($3H$) is not modified, in contrast to the generalized redshift term $mH$ for curvature perturbations. Consequently, their spectral indices, $n_T$ and $n_s-1$, can be significantly different. In Fourier space, the equation of motion for the tensor mode is given by:
\begin{equation}
H_r^{\prime\prime}(\mathbf{k},\eta)+\left(k^2-\frac{a^{\prime\prime}}{a}\right)H_r(\mathbf{k},\eta)=0~, \label{eq:hreom}
\end{equation}
where $H_r(\mathbf{k},\eta)\equiv a(\eta)h_r(\mathbf{k},\eta)$, $h_r$ denotes the tensor part of the metric perturbation ($h_+$ and $h_\times$), and the superscript $^\prime$ denotes derivatives with respect to the conformal time $\eta$, $d\eta=dt/a(t)$. 

Solving Eq.~(\ref{eq:hreom}) with $a(\eta)\propto \eta^\nu$ for the leading order term in the limit $k\eta\rightarrow 0$, one obtains the power spectrum of tensor modes,
\begin{equation}
\mathcal{P}_{h}\equiv \frac{k^3|h_r|^2}{2\pi^2}\propto k^{-|2\nu-1|+3}\eta^{-[2\nu-1+|2\nu-1|]}~.
\end{equation}
The spectral index of tensor modes ($\mathcal{P}_{h}\propto k^{n_T}$) is measured when they re-enter the horizon ($k\eta=1$) at the pivot scale $k_\ast=0.05~\mathrm{Mpc}$, which leads to
\begin{equation}\label{eq:smpshnt}
\mathcal{P}_{h}\propto k^{-|2\nu-1|+3}\left(k^{-1}\right)^{-[2\nu-1+|2\nu-1|]} \propto k^{2\nu+2}~,
\end{equation}
and reproduces $n_T=2\nu+2$. Using $a(\eta)\propto \eta^{\frac{2}{3w+1}}$ obtained by solving the Friedmann equation, which indicates $\nu=\frac{2}{3w+1}$, we re-obtain $n_T = \frac{4}{3w+1} + 2$. Using Eq.~(\ref{eq:smmnurest}) and Eq.~(\ref{eq:smpshnt}), we reproduce the well-known result for standard inflation ($(m,\nu)=(3,-1)$), $n_s-1=0$ and $n_T=0$, in this dual scenario description.

\section{\label{smsec:modelprior}SGWB spectrum, Prior, and Posterior Distributions of Each Model}
This section provides a summary of the stochastic gravitational wave background (SGWB) spectrum and the prior and posterior distributions for each model discussed in the main text.
\begin{enumerate}
    \item Model-1: The dual scenario of generalized inflation and bounce cosmologies in terms of the parameter space $(n_T, r)$, specifically described by Eq.(2) in the main paper. The SGWB spectrum of this model is given by:
    \begin{equation}
    \Omega_\mathrm{GW}(f)h^2 = \frac{3}{128} \Omega_{\gamma 0} h^2 \cdot r \cdot \mathcal{P}_\mathcal{R} \left(\frac{f}{f_\ast}\right)^{n_T} \left[\frac{1}{2} \left(\frac{f_\mathrm{eq}}{f}\right)^2 + \frac{16}{9}\right]~.
    \end{equation}
    Here, the prior parameters, $r$ and $n_T$, denote, respectively, the tensor-to-scalar ratio ($r \equiv \mathcal{P}_h/\mathcal{P}_\mathcal{R}$) and the spectral index of the primordial tensor spectrum at the pivot scale $k_\ast = 0.05 \mathrm{Mpc}^{-1}$. $\mathcal{P}_\mathcal{R} = 2 \times 10^{-9}$ is the amplitude of the curvature perturbation spectrum at $k_\ast$ \cite{Caprini:2018mtu}. Additionally, $f_\ast = 1.55 \times 10^{-15} \mathrm{Hz}$ corresponds to the frequency today for $k_\ast$, while $f_\mathrm{eq} = 2.01 \times 10^{-17} \mathrm{Hz}$ represents the frequency today at matter-radiation equality. $\Omega_{\gamma 0} = 2.474 \times 10^{-5} h^{-2}$ is the energy density fraction of radiation today, where $h = 0.6770$ is the reduced Hubble constant.
    
    \item Model-2: The dual scenario of generalized inflation and bounce cosmologies in terms of the parameter space $(w,r)$, utilizing the equation $n_T = \frac{4}{3w+1} + 2$ for $w\in(-\infty,\infty)$, specifically described by Eq.(3) in the main paper. The SGWB spectrum of this model is given by:
    \begin{equation}
    \Omega_\mathrm{GW}(f)h^2=\frac{3}{128}\Omega_{\gamma 0}h^2\cdot r\cdot\mathcal{P}_\mathcal{R}\left(\frac{f}{f_\ast}\right)^{\left[\frac{4}{3w+1}+2\right]}\left[\frac{1}{2}\left(\frac{f_\mathrm{eq}}{f}\right)^2+\frac{16}{9}\right].
    \end{equation}
    
    \item Model-3: The time-independent (stable) scale-invariant solution of the dual scenario in terms of parameter space $(m,r)$, utilizing the $m=-\frac{2}{\nu}+1$ for $m\in(-\infty,\infty)$, specifically described by Eq.(5) in the main paper. The SGWB spectrum of this model is given by:
    \begin{equation}\label{eq:smm3}
    \Omega_\mathrm{GW}(f)h^2=\frac{3}{128}\Omega_{\gamma 0}h^2\cdot r\cdot\mathcal{P}_\mathcal{R}\left(\frac{f}{f_\ast}\right)^{\left[-\frac{4}{m-1}+2\right]}\left[\frac{1}{2}\left(\frac{f_\mathrm{eq}}{f}\right)^2+\frac{16}{9}\right]~.
    \end{equation}
    
    \item Model-4: The time-dependent (dynamic) scale-invariant solution of the dual scenario in terms of parameter space $(m,r)$, utilizing the $m=\frac{4}{\nu}+1$ for $m\in(-\infty,\infty)$, specifically described by Eq.(6) in the main paper. The SGWB spectrum of this model is given by:
    \begin{equation}\label{eq:smm4}
    \Omega_\mathrm{GW}(f)h^2=\frac{3}{128}\Omega_{\gamma 0}h^2\cdot r\cdot\mathcal{P}_\mathcal{R}\left(\frac{f}{f_\ast}\right)^{\left[\frac{8}{m-1}+2\right]}\left[\frac{1}{2}\left(\frac{f_\mathrm{eq}}{f}\right)^2+\frac{16}{9}\right]~.
    \end{equation}
    
    \item Model-5: The time-independent (stable) nearly scale-invariant solution of the dual scenario in terms of parameter space $(m,r)$, utilizing the $m=-\frac{2-(n_s-1)}{\nu}+1$ for $m\in(-\infty,\infty)$ and $n_s-1=-0.04$. The SGWB spectrum of this model is given by:
    \begin{equation}\label{eq:smm5}
    \Omega_\mathrm{GW}(f)h^2=\frac{3}{128}\Omega_{\gamma 0}h^2\cdot r\cdot\mathcal{P}_\mathcal{R}\left(\frac{f}{f_\ast}\right)^{\left[-\frac{4.08}{m-1}+2\right]}\left[\frac{1}{2}\left(\frac{f_\mathrm{eq}}{f}\right)^2+\frac{16}{9}\right]~.
    \end{equation}
    
    \item Model-6: The time-dependent (dynamic) nearly scale-invariant solution of the dual scenario in terms of parameter space $(m,r)$, utilizing the $m=\frac{4+(n_s-1)}{\nu}+1$ for $m\in(-\infty,\infty)$ and $n_s-1=-0.04$. The SGWB spectrum of this model is given by:
    \begin{equation}\label{eq:smm6}
    \Omega_\mathrm{GW}(f)h^2=\frac{3}{128}\Omega_{\gamma 0}h^2\cdot r\cdot\mathcal{P}_\mathcal{R}\left(\frac{f}{f_\ast}\right)^{\left[\frac{7.92}{m-1}+2\right]}\left[\frac{1}{2}\left(\frac{f_\mathrm{eq}}{f}\right)^2+\frac{16}{9}\right]~.
    \end{equation}
    
    \item Model-7: Supermassive black hole binaries (SMBH binaries). The SGWB spectrum of this model is given by~\cite{Hobbs:2009yn, Bian:2023dnv}:
    \begin{equation}
    \Omega_{\mathrm{GW}}^{\mathrm{SMBHB}}(f) h^2=A_{\mathrm{SMBHB}}^2 \frac{2 \pi^2}{3 H_0^2} f^{5-\gamma} f_\mathrm{yr}^{\gamma-3} h^2.
    \end{equation}
    Here $\gamma=13/3$ is taken for SMBH binaries\cite{Jaffe:2002rt,Wyithe:2002ep}, $A_{\mathrm{SMBHB}}$ is the amplitude of SGWB spectrum, and $f_\mathrm{yr}\equiv 1/\mathrm{year}$. Note that our result is consistent with \cite{Figueroa:2023zhu}, where a different spectrum form for SMBH binaries is adopted.

    \item Model-8: Cosmic strings (CS). The SGWB spectrum of this model is given by~\cite{NANOGrav:2023hvm} (see references therein):
    \begin{equation}
    \Omega_{\mathrm{GW}}^{\mathrm{CS}}(f)h^2=\frac{8 \pi}{3 H_0^2}(G \mu)^2 \sum_{k=1}^{k_{\max }} P_k \mathcal{I}_k(f)h^2~.
    \end{equation}
    Following~\cite{NANOGrav:2023hvm}, the prior parameter $G\mu$ denotes cosmic string tension in units of Newton's constant. $P_k$ represents the gravitational wave power emitted by a loop in its $k$-th excitation,
    \begin{equation}
    P_k = \frac{\Gamma}{\zeta(q)} \frac{1}{k^q},
    \end{equation}
    where $\zeta(q)$ is determined by $\sum_k P_k = \Gamma$. $\Gamma = 50$ \cite{Blanco-Pillado:2017oxo} and $q = 4 / 3$ \cite{Chang:2021afa} are applicable to the case: stable strings emitting GWs only as bursts from cusps on closed loops~\cite{NANOGrav:2023hvm}. The frequency-dependent factor $\mathcal{I}_k(f)$ is given by:
    \begin{equation}
    \mathcal{I}_k(f) = \frac{2k}{f} \int_{t_{\text{ini}}}^{t_0} dt \left(\frac{a(t)}{a(t_0)}\right)^5 n_l\left(\frac{2k}{f} \frac{a(t)}{a(t_0)}, t\right),
    \end{equation}
    where $t_0$ is the present time, and $t_{\text{ini}}$ is when the network reaches the scaling attractor solution. The loop number density $n_l$ is estimated using the velocity-dependent one-scale (VOS) model:
    \begin{equation}
    n_l(\ell, t) = \mathcal{F} \frac{C_* \Theta(t - t_*) \Theta(t_* - t_{\text{ini}})}{\alpha_* (\alpha_* + \Gamma G \mu + \dot{\alpha}_* t_*) t_*^4} \left(\frac{a_*}{a(t)}\right)^3,
    \end{equation}
    with $\mathcal{F} = 0.1$ \cite{Blanco-Pillado:2013qja, Auclair:2019wcv} being an efficiency factor. An asterisk indicates evaluation at loop formation time $t_*$, obtained by solving:
    \begin{equation}
    t_* = \frac{\ell + \Gamma G \mu t}{\alpha_* + \Gamma G \mu}, \quad \alpha_* = \alpha(t_*).
    \end{equation}
    The time-dependent functions $C$ and $\alpha$ describe loop formation efficiency and typical loop size at production time:
    \begin{equation}
    C(t) = \frac{\tilde{c}}{\sqrt{2}} \frac{\bar{v}(t)}{\xi^3(t)}, \quad \alpha(t) = \alpha_L \xi(t).
    \end{equation}
    Here, $\tilde{c} \approx 0.23$ and $\alpha_L \approx 0.37$ agree with numerical simulations \cite{Blanco-Pillado:2013qja, Martins:2000cs, Blanco-Pillado:2011egf}. At high temperatures, $\xi_r \approx 0.27$ and $\bar{v}_r \approx 0.66$, leading to $\alpha \approx 0.10$ in the radiation-dominated era, aligns well with numerical simulations \cite{Blanco-Pillado:2013qja}.  

    \item Model-9: Domain walls (DW). The SGWB spectrum of this model is given by \cite{Chang:2021afa,Kadota:2015dza,Zhou:2020ojf}(see references therein):
    \begin{equation}
    \Omega_{\mathrm{GW}}^{\mathrm{DW}}(f)h^2=\Omega_{\mathrm{GW}}^{\mathrm{peak}} h^2 S^{\mathrm{DW}}(f)
    \end{equation}
    where the peak GW amplitude is
    \begin{equation}
    \begin{aligned}
    \Omega_{\mathrm{GW}}^{\text{peak}} h^2 \simeq  5.20 \times 10^{-20} \times \tilde{\epsilon}_{\mathrm{gw}} \mathcal{A}^4\left(\frac{10.75}{g_*}\right)^{1 / 3} \times \left(\frac{\sigma}{1 \mathrm{TeV}^3}\right)^4\left(\frac{1 \mathrm{MeV}^4}{\Delta V}\right)^2
    \end{aligned}
    \end{equation}
    and the shape function $S^{\mathrm{DW}}(f)$ is
    \begin{equation}
    \begin{aligned}
    S^{\mathrm{DW}}(f)=\left(f / f_{\text{peak}}^{\text{DW}}\right)^3, \quad f<f_{\text{peak}}^{\text{DW}} \\
    S^{\mathrm{DW}}(f)=\left(f / f_{\text{peak}}^{\text{DW}}\right)^{-1}, \quad f \geq f_{\text{peak}}^{\text{DW}}
    \end{aligned}
    \end{equation}
    where the peak frequency is \cite{Chang:2021afa}
    \begin{equation}
    f_{\text{peak}}^{\mathrm{DW}} \simeq 3.99 \times 10^{-9} \mathrm{Hz} \mathcal{A}^{-\frac{1}{2}}\left(\frac{1 \mathrm{TeV}^3}{\sigma}\right)^{\frac{1}{2}}\left(\frac{\Delta V}{1 \mathrm{MeV}^4}\right)^{\frac{1}{2}}
    \end{equation}
    Here, the prior parameters $\sigma$ and $\Delta V$ respectively represent the tension of the domain wall and the bias potential that breaks the degeneracy of the vacua. The bias potential leads to the disappearance of the domain wall and marks the location of the peak frequency. $\mathcal{A}=1.2$ \cite{Chang:2021afa, Kadota:2015dza} is the area parameter. $\tilde{\epsilon}_{\mathrm{gw}}=0.7$ \cite{Chang:2021afa} is the efficiency parameter for generating gravitational waves.
    
    \item Model-10: First-order phase transitions (FOPT). The SGWB spectrum of this model is given by \cite{Bian:2023dnv, Caprini:2015zlo}(see references therein):
    \begin{equation}
    \begin{aligned}
    \Omega_{\mathrm{GW}}^{\mathrm{FOPT}}(f) h^2 
    & =2.65 \times 10^{-6}\left(H_* \tau_{\mathrm{sw}}\right)\left(\frac{\beta}{H_*}\right)^{-1} v_b\left(\frac{\kappa_v \alpha_{PT}}{1+\alpha_{PT}}\right)^2 \\
    & \times \left(\frac{g_*}{100}\right)^{-\frac{1}{3}}\left(\frac{f}{f_{\text{peak}}^{\text{sw}}}\right)^3\left[\frac{7}{4+3\left(f / f_{\text{peak}}^{\text{sw}}\right)^2}\right]^{7 / 2},
    \end{aligned}
    \end{equation}
    with the peak frequency
    \begin{equation}
    f_{\text{peak}}^{\mathrm{FOPT}}=1.9 \times 10^{-5} \frac{\beta}{H_*} \frac{1}{v_b} \frac{T_*}{100}\left(\frac{g_*}{100}\right)^{\frac{1}{6}} \mathrm{Hz}.
    \end{equation}
    Here, $\tau_{\mathrm{sw}}=\min \left[\frac{1}{H_*},\frac{R_s}{U_f}\right]$ denotes the duration of the sound wave phase. $H_*$ is the Hubble parameter at temperature $T_*$. The proportion of vacuum energy allocated to each source, denoted as $\kappa_{\nu}$ \cite{Hirose:2023bvl, Zhou:2020ojf}, is related to $\alpha_{PT}$, which represents the latent heat. Additionally, $\beta / H_*$ corresponds to the inverse duration of the phase transition, $v_b$ denotes the velocity of the vacuum bubble wall in the plasma background. $g_*$ is the effective number of degrees of freedom for relativistic particles in the cosmic plasma at the time of gravitational wave formation.
    
    \item Model-11: Scalar-induced gravitational waves (SIGW). The SGWB spectrum of this model is given by \cite{Kohri:2018awv}(see references therein):
    \begin{equation}
    \begin{aligned}
    \Omega_{\mathrm{GW}}^{\mathrm{SI}}(f) h^2 & =\frac{1}{12} \Omega_{\mathrm{rad}} h^2\left(\frac{g_0}{g_*}\right)^{\frac{1}{3}} \times \int_0^{\infty} d v \int_{|1-v|}^{1+v} d u\left(\frac{4 v^2-\left(1+v^2-u^2\right)^2}{4 u v}\right)^2 \\
    & \times P_{\mathcal{R}}(2 \pi f u) P_{\mathcal{R}}(2 \pi f v) I^2(u, v),
    \end{aligned}
    \end{equation}
    where
    \begin{equation}
    \begin{aligned}
    I^2(u, v) & =\frac{1}{2}\left(\frac{3}{4 u^3 v^3 x}\right)^2\left(u^2+v^2-3\right)^2 \\
    & \times \left\{\left[-4 u v+\left(u^2+v^2-3\right) \ln \left|\frac{3-(u+v)^2}{3-(u-v)^2}\right|\right]^2\right. \\
    & \left.+\left[\pi\left(u^2+v^2-3\right) \Theta(u+v-\sqrt{3})\right]^2\right\} .
    \end{aligned}
    \end{equation}
    Here, $P_{\mathcal{R}}(k)$ represents the power spectrum of curvature perturbations. We consider a scenario where $P_{\mathcal{R}}(k)$ adopts a power-law form near $k = 2 \pi f_\mathrm{yr}=20.6 \mathrm{pc}^{-1}$:
    \begin{equation}
    P_{\mathcal{R}}(k)=P_{\mathcal{R} 0}\left(\frac{k}{k_*}\right)^n \Theta\left(k-k_{\min}\right) \Theta\left(k_{\max}-k\right),
    \end{equation}
    where $\Theta(x)$ is the Heaviside function, $P_{\mathcal{R} 0}$ is the power spectrum of scalar perturbations at $k = k_*$, $n$ represents the slope of the power spectrum for scalar perturbations. To prevent an excessive production of primordial black holes (PBHs), cutoffs are set at $k_{\min} = 0.03 k_*$ and $k_{\max} = 100 k_*$, imposing an upper limit on $P_{\mathcal{R}}(k)$ \cite{Bian:2023dnv}.
\end{enumerate}

In Table.\ref{Tab:prior}, we assign uniform or log-uniform priors to all parameters for each model, which are utilized for Bayesian analysis in this work.

\begin{table}[htbp]
\hspace{-1cm}
\centering
\caption{Summary of all priors}
\label{Tab:prior}
\begin{tabular}{ccc}
\toprule[0.4mm]
\textbf{Parameter} & \textbf{Description} & \textbf{Prior} \\ \hline
 & \textbf{Model-1: Dual scenario $(n_T, r)$} &  \\
$n_T$ & Spectral index of tensor spectrum & uniform $[-1, 6]$ \\
$r$ & Tensor-to-scalar ratio & log-uniform $[-16, 0]$ \\ \hline

 & \textbf{Model-2: Dual scenario $(w, r)$} &  \\
$w$ & Equation of state & uniform $[-10, 10]$ \\
$r$ & Tensor-to-scalar ratio & log-uniform $[-16, 0]$ \\\hline

 & \textbf{Model-3: Stable Scale-invariant $(m,r)$} &  \\
$m$ & Stable scale-invariant factor & uniform $[-32, 32]$ \\
$r$ & Tensor-to-scalar ratio & log-uniform $[-16, 0]$ \\ \hline

 & \textbf{Model-4: Dynamic Scale-invariant $(m,r)$} & \\
$m$ & Dynamic scale-invariant factor & uniform $[-32, 32]$ \\
$r$ & Tensor-to-scalar ratio & log-uniform $[-16, 0]$ \\ \hline

 & \textbf{Model-5: Stable nearly Scale-invariant $(m,r)$} & \\
$m$ & Stable nearly scale-invariant factor & uniform $[-32, 32]$ \\
$r$ & Tensor-to-scalar ratio & log-uniform $[-16, 0]$ \\ \hline

 & \textbf{Model-6: Dynamic nearly Scale-invariant $(m,r)$} & \\
$m$ & Dynamic nearly scale-invariant factor & uniform $[-32, 32]$ \\
$r$ & Tensor-to-scalar ratio & log-uniform $[-16, 0]$ \\ \hline

 & \textbf{Model-7:SMBHBs (PL $n_t = 13/3$)} & \\
$A_\mathrm{SMBHB}$ & Amplitude of the signal & log-uniform $[-18, -12]$\\ \hline

 & \textbf{Model-8:CS} & \\
$G\mu$ & Cosmic-string tension & log-uniform $[-12, -6]$\\ \hline

 & \textbf{Model-9:DW} & \\
$\sigma$ & Surface energy density & log-uniform $[0, 8]$\\ 
$\Delta V$ & Bias potential & log-uniform $[0, 8]$\\ \hline

 & \textbf{Model-10:FOPT} & \\
$\beta /H_\star$ & Inverse PT duration & uniform $[5, 70]$\\ 
$T_\star$ & PT temperature & uniform $[0.01, 1.6]$\\
$\alpha_{PT}$ & PT strength & uniform $[0.0, 1.0]$\\ 
$\nu_b$ & Velocity of bubble wall & uniform $[0.9, 1.0]$\\ \hline

 & \textbf{Model-11: SIGW} & \\
$P_{R0}$ & Amplitude of signal & log-uniform $[-4, -1.5]$\\ 
$n$ & Slope of spectrum & uniform $[-2, 2]$\\ 
\toprule[0.4mm]
\end{tabular}
\end{table}

\section{Data analysis methodology and Bayes factor  matrices}

\begin{figure}[htbp]
\centering 
\includegraphics[width=1.0\textwidth]{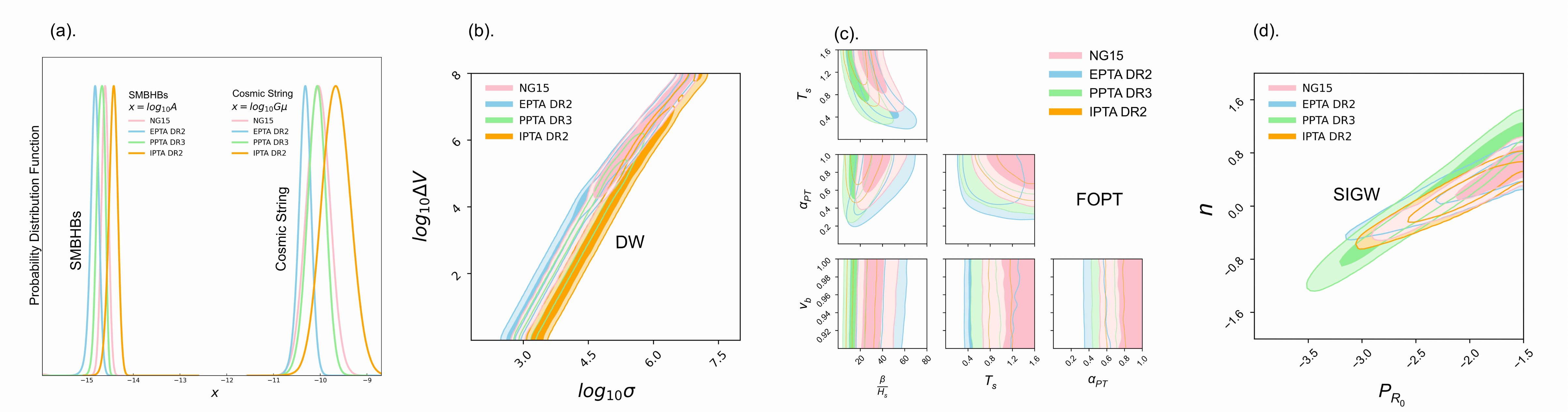}
\caption{\label{fig:Others.pdf} Probability distribution function and posterior distributions (at $68\%$ and $95\%$ confidence levels) for the conventional other astrophysical and cosmological source. (a): SMBH binaries and cosmic string; (b) Domain wall; (c) First order phase transition; and (d) Scalar-induced gravitational waves. They are consistent with previous results in \cite{Bian:2023dnv, NANOGrav:2023hvm}.}
\end{figure}

In our Bayesian analysis, we employed \texttt{Ceffyl}~\cite{Lamb:2023jls} to fit Model-1 to Model-11, as listed in last section, within their respective prior ranges specified in Table~\ref{Tab:prior}, using data from NANOGrav 15-year \cite{ZenodoNG, NANOGrav:2023hde, NANOGrav:2023gor, NANOGrav:2023hvm}, EPTA DR2 \cite{ZenodoEPTA, EPTA:2023sfo, EPTA:2023fyk}, PPTA DR3 \cite{PPTADR3, Zic:2023gta, Reardon:2023gzh}, and IPTA DR2 \cite{IPTADR2, Perera:2019sca, Antoniadis:2022pcn} datasets. Specifically, we utilized the \texttt{PTMCMC} sampler in \texttt{Ceffyl} to generate the posterior distribution for each case, as illustrated in Fig.2 in the main paper and Fig.~\ref{fig:Nearmrplot.pdf} for the dual scenario, and in Fig.~\ref{fig:Others.pdf} for other conventional models. We used \texttt{UltraNest} in \texttt{Ceffyl} to compute the Bayes Factors for each case, as listed in Tab.~\ref{tab:smnew1}-Tab.~\ref{tab:smnew3} and Eq.~(\ref{eq:smNG15bfm})-Eq.~(\ref{eq:smIPTAbfm}). In particular, in Eq.~(\ref{eq:smNG15bfm})-Eq.~(\ref{eq:smIPTAbfm}), the Bayes factor (BF) matrices, $B_{ij} = \mathrm{evidence}[H_i]/\mathrm{evidence}[H_j]$, are organized in the order $i,j=$ SMBHBs, CS, DW, FOPT, and SIGW. The Bayes factor matrices for the conventional sources are consistent with previous results in \cite{Bian:2023dnv, NANOGrav:2023hvm}. For CS and SIGW, we also utilized \texttt{PTArcade} \cite{Mitridate:2023oar}. To visualize the probability and posterior distributions, we used \texttt{Getdist} \cite{Lewis:2019xzd} and \texttt{Corner} \cite{Corner}. Following \cite{Bian:2023dnv}, we list the interpretation of the Bayes factor for model comparisons in Tab.~\ref{tab:sminterbf}: a Bayes factor $B_{ij}$ of 20 between a candidate model $\mathrm{M}_i$ and another model $\mathrm{M}_j$ corresponds to a 95\% belief in the statement ``$\mathrm{M}_i$ is true,'' indicating strong evidence in favor of $\mathrm{M}_i$ \cite{Bian:2023dnv}. Moreover, a Bayes factor $B_{ij}$ of 1 between a candidate model $\mathrm{M}_i$ and another model $\mathrm{M}_j$ indicates no preference between them \cite{Bian:2023dnv}.

\begin{table}[htbp]
\centering
\caption{Bayes Factor of Model-2: dual scenario \((w, r)\) versus other Models}\label{tab:smnew1}
\hspace*{-1cm}
\begin{tabular}{cccccc}
\toprule
 &SMBHBs & CS & DW & FOPT & SIGW  \\ \hline
NG15 & 94.6 ± 34.9 & 461.6 ± 127.2 & 318.8 ± 73.3 & 61.4 ± 11.6 & 65.2 ± 12.4  \\
EPTA DR2 & 3.9 ± 1.6 & 3.1 ± 1.2 & 30.5 ± 8.0 & 3.2 ± 0.7 & 6.0 ± 1.4  \\ 
PPTA DR3 & 22.3 ± 9.1 & 16.4 ± 7.1 & 19.9 ± 8.9 & 6.4 ± 2.4 & 12.1 ± 4.2  \\
IPTA DR2 & 3.8 ± 2.0 & 7.6 ± 3.4 & 7.3 ± 3.7 & 2.8 ± 1.1 & 13.9 ± 6.2 \\ \bottomrule
\end{tabular}
\end{table}

\begin{table}[htbp]
\centering
\caption{Bayes Factor of Model-3: stable solutions \((m, r)\) versus other Models} \label{tab:smnew2}
\hspace*{-1cm}
\begin{tabular}{cccccc}
\toprule
 & SMBHBs & CS & DW & FOPT & SIGW  \\ \midrule
NG15 & 81.5 ± 32.3 & 397.7 ± 117.8 & 274.7 ± 67.8 & 52.9 ± 10.7 & 56.2 ± 10.7  \\
EPTA DR2 & 3.0 ± 1.3 & 2.4 ± 0.9 & 23.6 ± 6.1 & 2.5 ± 0.6 & 4.7 ± 1.1  \\ 
PPTA DR3 & 15.1 ± 6.2 & 11.1 ± 4.8 & 13.5 ± 6.0 & 4.3 ± 1.6 & 8.2 ± 2.9  \\
IPTA DR2 & 3.3 ± 1.7 & 6.6 ± 2.9 & 6.3 ± 3.2 & 2.4 ± 1.0 & 12.0 ± 5.4 \\ \bottomrule
\end{tabular}
\end{table}

\begin{table}[htbp]
\centering
\caption{Bayes Factor of Model-4: dynamic solutions \((m, r)\) versus other Models}\label{tab:smnew3}
\hspace*{-1cm}
\begin{tabular}{cccccc}
\toprule
 & SMBHBs & CS & DW & FOPT & SIGW \\ \midrule
NG15 & 49.4 ± 17.3 & 240.7 ± 63.2 & 166.3 ± 36.4 & 32.0 ± 6.8 & 34.0 ± 6.5 \\
EPTA DR2 & 2.3 ± 0.9 & 1.8 ± 0.7 & 17.7 ± 4.4 & 1.9 ± 0.4 & 3.5 ± 0.8 \\ 
PPTA DR3 & 12.3 ± 5.0 & 9.0 ± 3.9 & 11.0 ± 4.9 & 3.5 ± 1.3 & 6.7 ± 2.3  \\
IPTA DR2 & 2.2 ± 1.1 & 4.4 ± 1.9 & 4.2 ± 2.2 & 1.6 ± 0.7 & 8.0 ± 3.6 \\ \bottomrule
\end{tabular}
\label{tab:matrix_first_row_new3}
\end{table}

\begin{equation}\label{eq:smNG15bfm}
B^{NG15}_{ij}= \begin{pmatrix}
 1.0 & 4.8 \pm 1.9 & 3.4 \pm 1.2 & 0.7 \pm 0.2 & 4.0 \pm 1.1\\
 0.2 \pm 0.1 & 1.0 & 0.7 \pm 0.2 & 0.1 \pm 0.03 & 0.1 \pm 0.03  \\
 0.3 \pm 0.1 & 1.4 \pm 0.4 & 1.0 & 0.2 \pm 0.04 & 0.2 \pm 0.04 \\
 1.5 \pm 0.5 & 7.4 \pm 1.8 & 5.3 \pm 1.0 & 1.0 & 1.1 \pm 0.2 \\
 1.4 \pm 0.4 & 7.0 \pm 1.5 & 4.9 \pm 0.9 & 0.9 \pm 0.2 & 1.0 \\
\end{pmatrix}
\end{equation}

\begin{equation}\label{eq:smEPTAbfm}
B^{EPTA}_{ij} = \begin{pmatrix}
1.0 & 0.8 \pm 0.4 & 7.8 \pm 3.1 & 0.8 \pm 0.3 & 1.5 \pm 0.6  \\
1.3 \pm 0.6 & 1.0 & 9.9 \pm 3.6 & 1.1 \pm 0.4 & 2.0 \pm 0.6  \\
 0.1 \pm 0.1 & 0.1 \pm 0.04 & 1.0 & 0.1 \pm 0.02 & 0.2 \pm 0.1 \\
 1.2 \pm 0.5 & 1.0 \pm 0.3 & 9.4 \pm 1.8 & 1.0 & 1.9 \pm 0.4 \\
 0.6 \pm 0.2 & 0.5 \pm 0.2 & 5.0 \pm 1.3 & 0.5 \pm 0.1 & 1.0 \\
\end{pmatrix}
\end{equation}

\begin{equation}\label{eq:smPPTAbfm}
B^{PPTA}_{ij} = \begin{pmatrix}
1.0 & 0.7 \pm 0.3 & 0.9 \pm 0.4 & 0.3 \pm 0.1 & 0.5 \pm 0.2 \\
 1.4 \pm 0.6 & 1.0 & 1.2 \pm 0.5 & 0.4 \pm 0.2 & 0.7 \pm 0.2 \\
 1.1 \pm 0.5 & 0.8 \pm 0.4 & 1.0 & 0.3 \pm 0.1 & 0.6 \pm 0.2 \\
3.5 \pm 1.3 & 2.6 \pm 1.0 & 3.1 \pm 1.2 & 1.0 & 1.9 \pm 0.5  \\
 1.8 \pm 0.7 & 1.4 \pm 0.5 & 1.6 \pm 0.7 & 0.5 \pm 0.2 & 1.0 \\
\end{pmatrix}
\end{equation}

\begin{equation}\label{eq:smIPTAbfm}
B^{IPTA}_{ij} = \begin{pmatrix}
 1.0 & 2.0 \pm 1.1 & 1.9 \pm 1.1 & 0.8 \pm 0.4 & 3.7 \pm 1.7\\
 0.5 \pm 0.3 & 1.0 & 1.0 \pm 0.5 & 0.4 \pm 0.1 & 1.8 \pm 0.7\\
 0.5 \pm 0.3 & 1.0 \pm 0.5 & 1.0 & 0.4 \pm 0.2 & 1.9 \pm 0.5  \\
 1.3 \pm 0.6 & 2.7 \pm 1.0 & 2.6 \pm 1.2 & 1.0 & 4.9 \pm 1.5 \\
 0.3 \pm 0.01 & 0.6 \pm 0.2 & 0.5 \pm 0.5 & 0.2 \pm 0.1 & 1.0 \\
\end{pmatrix}
\end{equation}

\begin{table}[htbp]
\centering
\caption{Interpretation of Bayes Factor for model comparisons: a Bayes factor $B_{ij}$ of 20 between a candidate model $\mathrm{M}_i$ and another model $\mathrm{M}_j$ corresponds to a 95\% belief in the statement ``$\mathrm{M}_i$ is true,'' indicating strong evidence in favor of $\mathrm{M}_i$ \cite{Bian:2023dnv}.}\label{tab:sminterbf}
\hspace*{-1cm}
\begin{tabular}{cc}
\toprule
$B_{ij}$ & Evidence in favor of $\mathrm{M}_i$ against $\mathrm{M}_j$   \\ \hline
$1-3$ & Weak  \\
$3-20$ & Positive\\ 
$20-150$ & Strong  \\
$\ge 150$ & Very strong\\ \bottomrule
\end{tabular}
\end{table}

\section{Nearly scale-invariant solutions ($n_s-1=-0.04$)}
\begin{figure}[htbp]
\centering 
\includegraphics[width=0.8\textwidth]{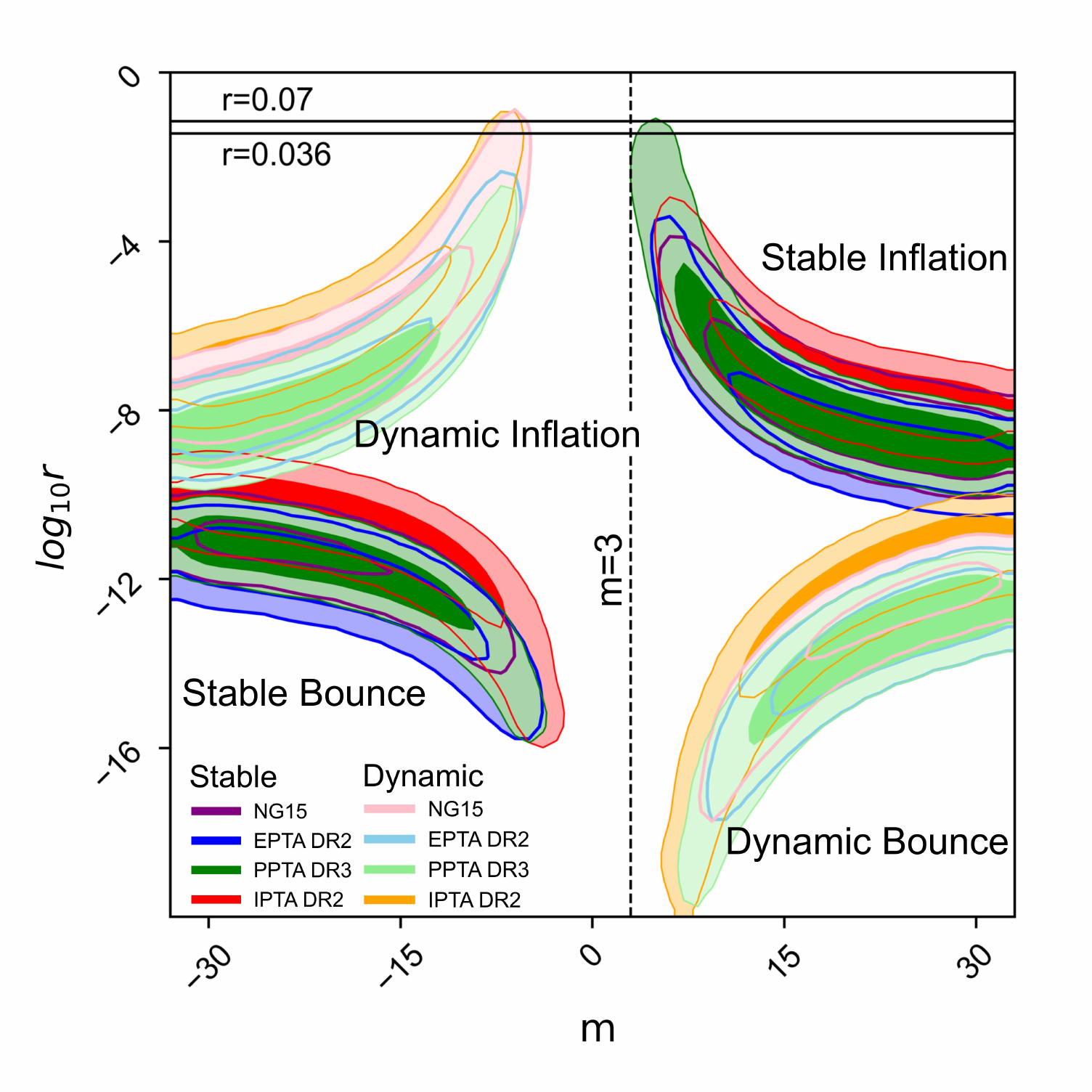}
\caption{\label{fig:Nearmrplot.pdf} The posterior distribution (at $68\%$ and $95\%$ confidence levels) in the parameter space $(m,r)$ of the nearly scale-invariant solutions (Eq.(\ref{eq:smm5}) and Eq.(\ref{eq:smm6})) of dual scenario of generalized inflation and bounce cosmologies, which closely resembles Fig.2(c) in the main paper, as expected.}
\end{figure}

For nearly scale-invariant solutions with $n_s-1=-0.04$, consistent with WMAP and Planck observations, their spectra (Eq.~(\ref{eq:smm5}) and Eq.~(\ref{eq:smm6})) exhibit slight differences from the purely scale-invariant solutions (Eq.~(\ref{eq:smm3}) and Eq.~(\ref{eq:smm4})) by a few percentage points. The posterior distribution for these solutions is illustrated in Fig.~\ref{fig:Nearmrplot.pdf}, which closely resembles Fig.2(c) in the main paper, as expected. This provides a cross-check for the results presented in the main text.

%% [A] Recommended: using JHEP.bst file
 \bibliographystyle{JHEP}
 \bibliography{biblio.bib}

%\begin{thebibliography}{99}
%\cite{Guth:1980zm}
%\bibitem{Guth:1980zm}
%A.~H.~Guth,
%``The Inflationary Universe: A Possible Solution to the Horizon and Flatness Problems,''
%Phys. Rev. D \textbf{23} (1981), 347-356
%doi:10.1103/PhysRevD.23.347
%9082 citations counted in INSPIRE as of 04 Sep 2022
%\end{thebibliography}

%%%%
%%%%

\end{document}